\documentclass[12pt]{article}
\usepackage[dvips]{graphicx}
\usepackage{epsfig,cite}
\usepackage {amsmath}
\usepackage{amssymb}

\newcommand\beq{\begin{equation}}
\newcommand\eeq{\end{equation}}
\newcommand\beqar{\begin{eqnarray}}
\newcommand\eeqar{\end{eqnarray}}
\def\fun#1#2{\lower3.6pt\vbox{\baselineskip0pt\lineskip.9pt
  \ialign{$\mathsurround=0pt#1\hfil##\hfil$\crcr#2\crcr\sim\crcr}}}

\begin{document}

\begin{titlepage}
\pagestyle{empty}
\begin{center}
{\large {\bf Managing Information for Sparsely Distributed Articles and Readers:  The Virtual Journals of the Joint Institute for Nuclear Astrophysics (JINA)}}
\end{center}

\begin{center}
{\bf Richard H. Cyburt}$^{1,2}$, {\bf Sam M. Austin}$^{1,2}$, {\bf Timothy C. Beers}$^{1,3,4}$, \\ {\bf Alfredo Estrade}$^{1,2,4}$, {\bf Ryan M. Ferguson}$^{1,2,4}$, {\bf A. Sakharuk}$^{1,2}$, {\bf Karl Smith}$^{1,2,4}$, \\ and {\bf Scott Warren}$^{1,2,5}$ \\
\end{center}
\vskip 0.1in
{\small {\it
$^1${Joint Institute for Nuclear Astrophysics (JINA), http://www.jinaweb.org} \\
$^2${National Superconduction Cyclotron Laboratory (NSCL), Michigan State University, East Lansing, MI 48824} \\
$^3${Center for the Study of Cosmic Evolution (CSCE), http://www.pa.msu.edu/astro/CSCE/} \\
$^4${Department of Physics and Astronomy, Michigan State University, East Lansing, MI 48824} \\
$^5${Department of Electrical and Computer Engineering, Michigan State University, East Lansing, MI 48824}} \\
}
\vskip 0.2in
\begin{center}
{\bf Abstract}
\end{center}
\baselineskip=18pt \noindent
{The research area of nuclear astrophysics is characterized by a need
for information published in tens of journals in several fields and an
extremely dilute distribution of researchers. For these reasons it is
difficult for researchers, especially students, to be adequately
informed of the relevant published research. For example, the commonly
employed journal club is inefficient for a group consisting of a
professor and his two students. In an attempt to address this problem,
we have developed a virtual journal (VJ), a process for collecting and
distributing a weekly compendium of articles of interest to
researchers in nuclear astrophysics.  Subscribers are notified of each
VJ issue using an email-list server or an RSS feed.  The VJ data base
is searchable by topics assigned by the editors, or by keywords. There
are two related VJs: the Virtual Journal of Nuclear Astrophysics (JINA
VJ), and the SEGUE Virtual Journal (SEGUE VJ). The JINA VJ also serves
as a source of new experimental and theoretical information for the
JINA REACLIB reaction rate database.  References to review articles
and popular level articles provide an introduction to the literature
for students.  The VJs and support information are available at
http://groups.nscl.msu.edu/jina/journals.}
\vfill
\leftline{[JINA] email: {\tt  cyburt@nscl.msu.edu, austin@nscl.msu.edu}}
\leftline{[SEGUE] email: {\tt beers@pa.msu.edu}}
\end{titlepage}
\baselineskip=18pt

\section{Introduction}
\label{sect:introduction}

Nuclear astrophysics is an interdisciplinary subject with
contributions from many branches of physics and astrophysics.  For
example, someone studying the r-process that synthesizes half of the
heavy elements needs familiarity with, for example, measurements of
nuclear lifetimes and nuclear masses, shell model calculations of
nuclear properties, models of supernova explosions, freeze-out
processes, abundances in metal poor stars, atomic oscillator
strengths, processes in stellar surfaces, and the implications of all
these for nucleocosmochronology. It is, therefore, difficult for
researchers in nuclear astrophysics, and especially for young
researchers, to be cognizant of all the literature bearing on their
research.  This is aggravated by the dilute distribution of
researchers, often only one in the field at a given institution.

To remedy this problem, in 2003 we founded and have since been
operating a Virtual Journal of Nuclear Astrophysics (JINA VJ).  It is
sponsored by JINA, the Joint Institute of Nuclear
Astrophysics~\cite{jina} at the National Superconducting Cyclotron
Laboratory (NSCL), Michigan State University.  In 2006, we founded a
similar journal for articles of particular interest to astronomical
observers, the SEGUE Virtual Journal (SEGUE VJ)\cite{segue}.
Essentially all the technical description below applies to both of
these virtual journals. More recently the JINA VJ has become an
important source of data for the on-line JINA REACLIB
database~\cite{reaclib}.

We describe in Section \ref{sect:architecture} the evolution of
procedures for generating the journal; in Section
\ref{sect:featuresfunctions} the features and functions of the JINA VJ
and some examples of its usage; and in Section \ref{sect:SVJ} the
extension of these procedures to the Segue VJ.

\section{Architecture}
\label{sect:architecture}

To provide the desired coverage it was necessary to scan for
appropriate articles in over forty journals, as shown in
Table~\ref{tab:journals}. It was found to be important to have two
people with broadly different expertise perform redundant scans to
ensure that the appropriate choices were made.  Initially this process
involved the following steps. (1) An undergraduate student provided
the Table of Contents (TOC's) of the chosen journals, including links to their
on-line versions. (2) These lists were scanned by two editors, a
nuclear physicist and an astronomer.  For each chosen paper, one or
more categories were assigned, for example, ``reaction rate'', ``r
process'', etc., to aid in searching by users.  The categories for the
JINA VJ are shown in Fig.~\ref{fig:search}(a).  (3) An undergraduate
student then compiled the chosen titles, links to them, and their
assigned categories into a list which was entered into a searchable
database maintained by the NSCL. A Listserv notification was then sent
to enrolled subscribers.

\begin{figure}[ht]
\begin{center}
\epsfig{file=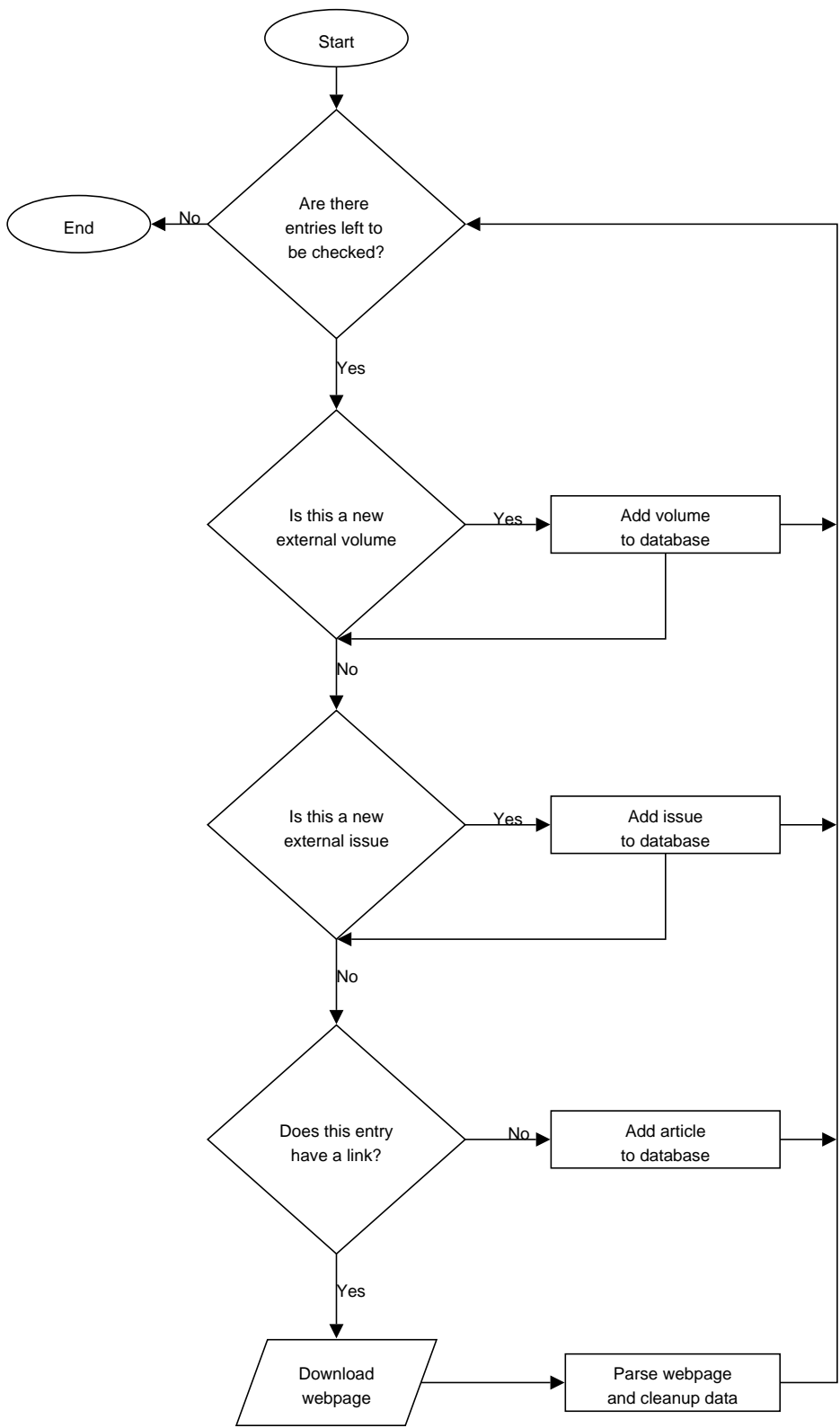,height=0.95\textheight}
\caption{\label{fig:download}
The article download process.}
\end{center}
\end{figure}

\begin{figure}[ht]
\begin{center}
\epsfig{file=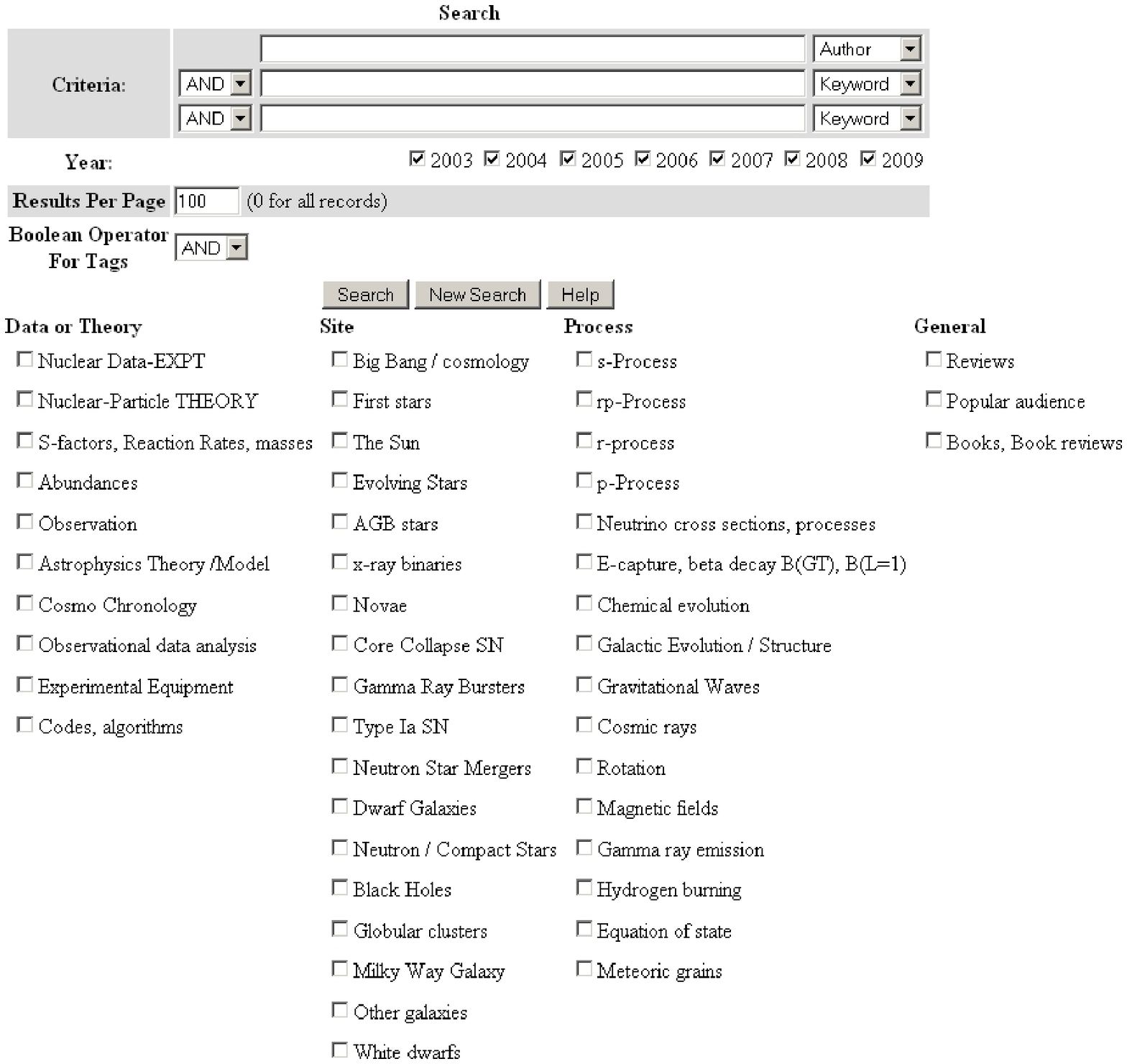,width=0.49\textwidth}
\epsfig{file=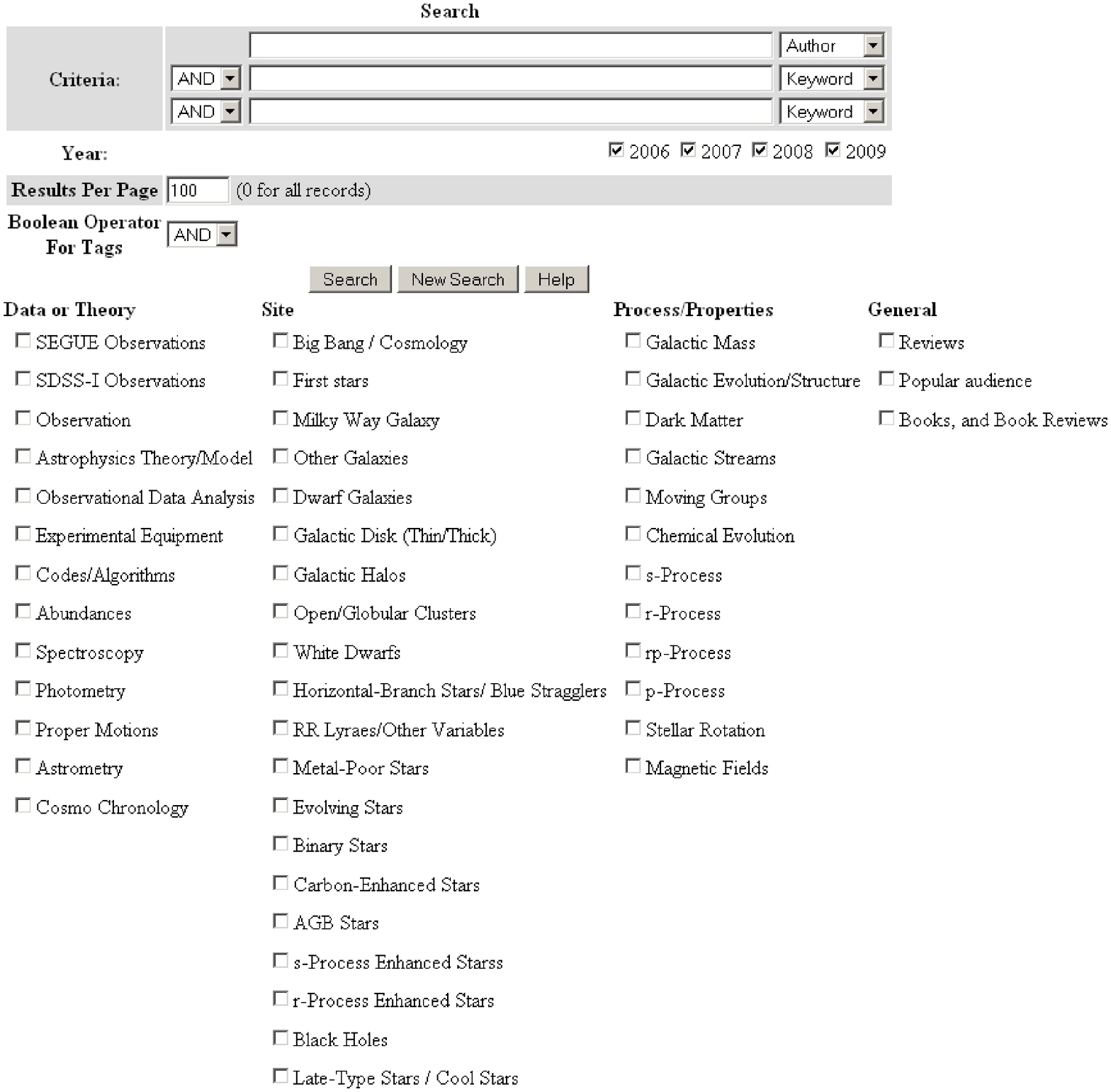,width=0.49\textwidth}
\caption{\label{fig:search}
The category choices available for searches within the
(a) JINA VJ (on left) and (b) SEGUE VJ (on right), by checking the small
boxes.  Both AND and OR searches are available.  They may be combined
with words entered as text in the three search boxes and choices of
years to be searched.}
\end{center}
\end{figure}

\begin{figure}[ht]
\begin{center}
\epsfig{file=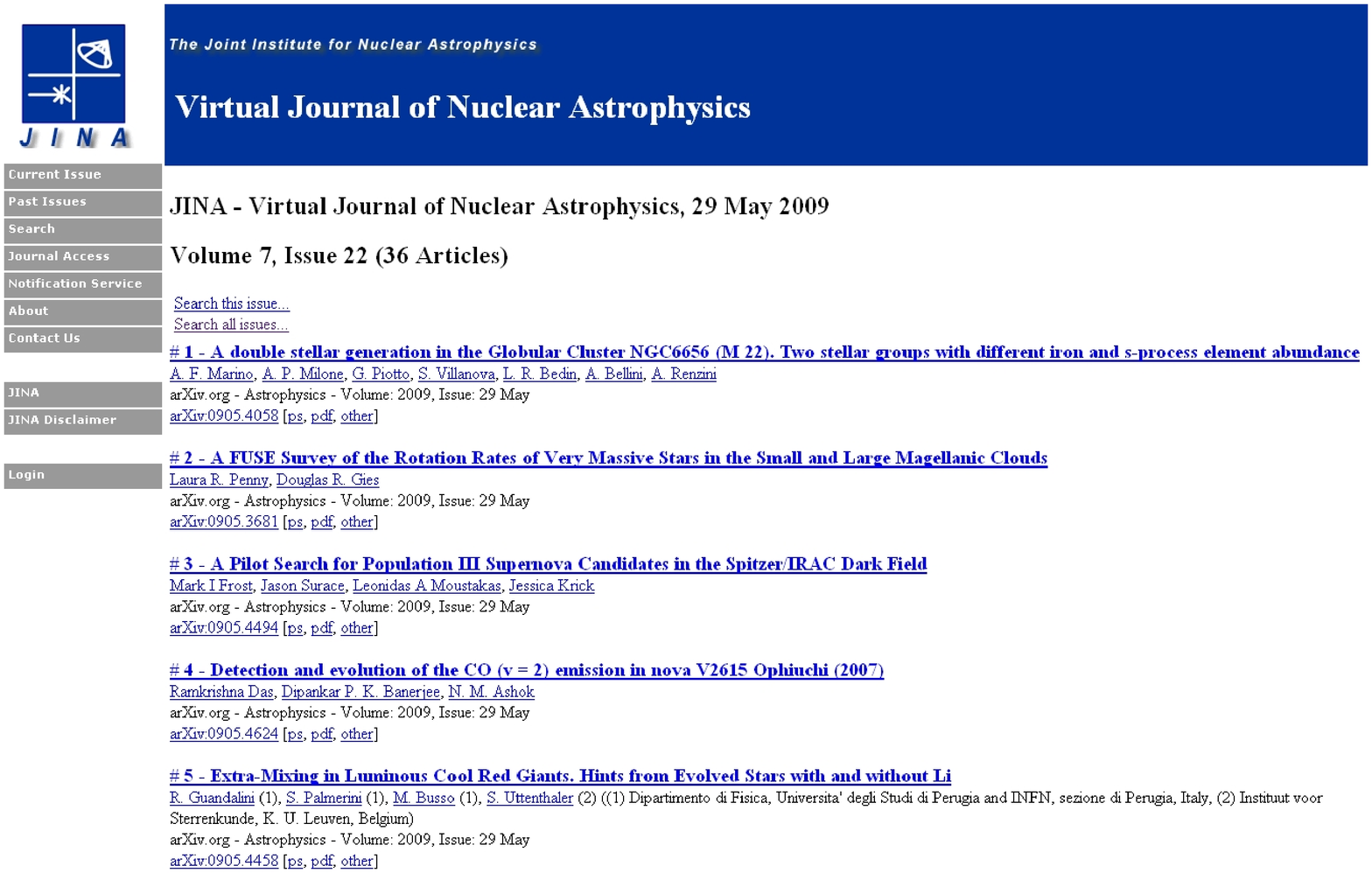,width=\textwidth}
\caption{\label{fig:page}
Part of a page from the virtual journal.}
\end{center}
\end{figure}

More recently these processes have been streamlined. Step~(1) The
production of the TOC's is now done semi-automatically.  It was
attempted to make this process completely automatic, but continual
structural changes in the journal TOC's made this untenable, and we
have reverted to a less automated process.  This is achieved with a
suite of regular expressions for each journal
%
%
.  As journal TOC's
change, these regular expressions can easily be updated.
Figure~\ref{fig:download} shows a schematic of the download process.
After the weekly downloads, any error messages from problem downloads
are examined and changes implemented.  This takes about 30 minutes per
week. Step~(2) The lists are made available on a web-based
PHP-driven~\cite{php} interface so they can be easily searched
remotely by the editors; this takes about 60 minutes per editor per
week. Step~(3) The list of titles, links, and categories, is entered
into a MySQL database~\cite{mysql} , and a notification is sent to
readers via "Listserv" or "RSS".  This takes about 10 minutes per
week.  An example of an issue of the JINA VJ is shown in
Fig.~\ref{fig:page}.  The MySQL-based database developed specifically
for the VJs provides flexible searches and integrates seamlessly with
the REACLIB
\cite{reaclib} database.  This database provides recommended nuclear 
reaction rates to the community of modelers of processes in the
cosmos, mostly in stars. More details about the virtual journal can be
found at: {\tt http://groups.nscl.msu.edu/jina/journals/jinavj}

\section{Features and functions}
\label{sect:featuresfunctions}

The first weekly issue of JINA VJ was published on July 18, 2003.  A
typical page is shown in Figure 2.  There are normally 20-60 articles
in a given issue and 1500 to 2000 articles in a given year; a total of
10381 articles have been included through the December 2008 issues.
The reader may click the appropriate link to see the abstracts or
complete articles for those publications that make them publicly
available without subscription.  Of course, access to complete
articles is, in many cases, available only if the reader's institution
has an appropriate subscription.

The individual journals contribute widely different numbers of
articles, with the preprint servers having the largest contributions.
To give an indication of the important journals for this field,
Table~\ref{tab:journals} notes the number of articles from each
journal for 2008.

Figure~\ref{fig:search}(a) shows the search page for the JINA VJ.
Checking any number of boxes and then performing a search presents a
list of articles labeled by all of the box-items (an AND search) or
any one of them (an OR search).  In addition, text can be entered in
the search boxes and ANDed or ORed with the other choices.  One can
limit the search to a particular year or years. This capability allows
an easy search of the literature.  It has also proven useful for
teaching classes in the field. For example, a choice of ``Reviews''
and ``Core collapse SN'' yields review articles on supernovae.  A
search on Woosley provides links to papers by S.E.~Woosley

\begin{figure}[ht]
\begin{center}
\epsfig{file=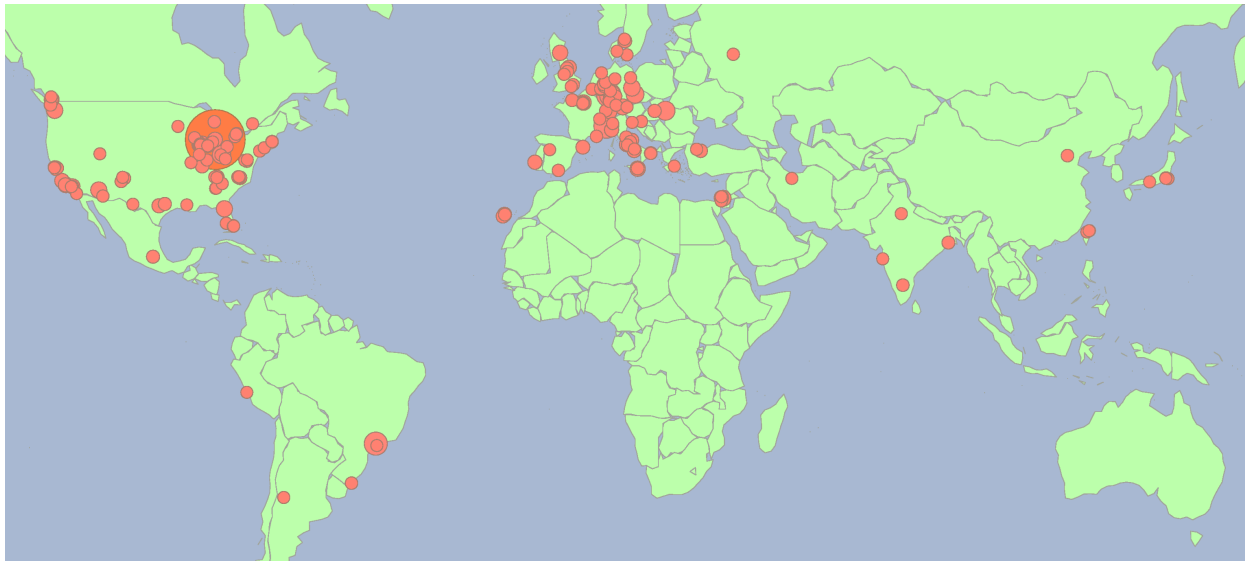,width=\textwidth}
\caption{\label{fig:map}
This map shows the cities from which subscribers are viewing the JINA VJ website, as determined from Google Analytics~\cite{google} over the last year.}
\end{center}
\end{figure}

The usage of the JINA VJ is quantitatively evaluated in two ways.
First, we keep track of the number of individuals who continue to
receive our weekly list server notification.  The list was initially
populated by large mailing lists for the field of nuclear
astrophysics; individuals can enlist or de-list easily.  The total
number notified has grown slowly. It is presently about 270; we find
that few de-list, except for young researchers who leave the field.
The second approach is through web based hit and visit analyzers.
Because of their contamination by commercial search firms and their
robots, these data are very hard to interpret.  None-the-less, for the
usual comparisons, we typically find 30000/5000 hits/visits per month.
A more reliable approach is through use of an engine such as Google
Analytics~\cite{google}, which is mainly insensitive to robot
searches.  The results since we have been using Analytics (April/2007
to May/2009) include 11863 visits, and $\sim$3 pages viewed per visit,
compared to the group of $\sim$270 individuals that receives weekly
notifications. Analytics also provides geographical distribution
statistics. Of the 11863 visitors 6640 are from the Americas, 4390
from Europe, 736 from Asia, 84 from Oceana and 8 from Africa. The
first ten countries, in order of visitors, are: United States,
Germany, Italy, Brazil, United Kingdom, France, Canada, Hungary,
Israel and Spain.  Shown in Figure~\ref{fig:map}, are the cities
viewing the VJ over the last year.  The wide geographical distribution
of viewers is a reflection of the access that the JINA VJ provides for
the sparse distribution of researchers in less developed countries.

\section{The SEGUE Virtual Journal}
\label{sect:SVJ}

During the evolution of JINA, astronomical observation came to play a
more and more important role.  It appeared that this observational
community would benefit from an additional virtual journal that would
provide coverage for observers that was inappropriate for the more
diverse clientele of the JINA VJ.  The SEGUE VJ was formed to serve
this need~\cite{segue}.  The operation of the SEGUE VJ takes advantage
of the procedures developed for JINA VJ.  The categories for SEGUE VJ
are shown in Fig.~\ref{fig:search}(b).  More details can be found at:
{\tt http://groups.nscl.msu.edu/jina/journals/seguevj}

\section{Conclusions}
\label{sect:conclusions}
We have developed an efficient set of procedures for producing Virtual
Journals: summaries of titles and links to articles from a set of
journals that are of interest to a particular research field. There
are convenient search procedures for users, including choices of
topics, key words, and authors.  It is possible to link seamlessly to
other databases that use material from the surveyed journals. The
database structure and interface routines and other software needs can
be made available, though with limited support.

\section{Acknowledgments}
We thank Jason Danielson for his assistance during the early
development of the JINA VJ.  This work is supported by the
U.S. National Science Foundation Grants No. PHY-02-016783 and
PHY-08-22648 (JINA).

\begin{table}[ht]
\begin{center}\caption{\label{tab:journals}
This table lists the journals scanned for the JINA VJ. The start date
for which the specified journal was scanned and the number
of articles from each journal that was chosen for inclusion in JINA VJ
in 2008 is shown. }
\begin{tabular}{||l|r|r||}
\hline\hline
Journal & 	Checked After & \# articles 2008 \\ \hline
American Journal of Physics &	2003-09-12 & 12 \\
Annual Review of Astronomy and Astrophysics &	2003-07-18 & 2 \\
Annual Review of Nuclear and Particle Science&	2003-07-18 & 1 \\
arXiv.org - Astrophysics&	2003-07-18 & 436 \\
arXiv.org - Nuclear Experiment&	2003-08-08 & 80 \\
arXiv.org - Nuclear Theory&	2003-07-18 & 108 \\
Astronomical Journal&	2005-09-30 & 23 \\
Astronomische Nachrichten&	2006-05-19 & 5 \\
Astronomy and Astrophysics&	2003-07-18 & 132 \\
Astronomy and Astrophysics Review&	2006-05-05 & 5 \\
Astronomy Letters&	2003-08-01 & 11 \\
Astronomy Reports&	2003-07-25 & 7 \\
Astroparticle Physics&	2004-07-23 & 2 \\
The Astrophysical Journal&	2003-07-18 & 113  \\
The Astrophysical Journal Letters&	2003-07-25 & 50 \\
The Astrophysical Journal Supplement&	2003-07-25 & 10 \\
Atomic Data and Nuclear Data Tables&	2005-01-06 & 1 \\
The European Physical Journal A&	2004-04-16 & 22 \\
JETP Letters&	2003-10-03 & 0 \\
Journal of Physics G&	2004-03-12 & 9 \\
Monthly Notices of the Royal Astronomical Society&	2003-08-15 & 67 \\
Nature&	2003-07-18 & 41 \\
Nature-Physics&	2006-02-10 & 1 \\
New Astronomy&	2003-08-29 & 3 \\
New Astronomy Reviews&	2004-09-10 & 9 \\
Nuclear Instr. and Methods in Phys. Research Sect. A&	2003-11-21 & 7 \\
Nuclear Physics A&	2003-07-25 & 15 \\
Nuclear Physics B&	2003-08-29 & 0 \\
Nuclear Physics News&	2006-07-21 & 4 \\
Physical Review C&	2003-07-18 & 126 \\
Physical Review D&	2003-07-18 & 26 \\
Physical Review Letters&	2003-07-18 & 35 \\
Physics Letters B&	2003-07-25 & 9 \\
Physics of Atomic Nuclei&	2003-07-18 & 0 \\
Physics of Particles and Nuclei&	2003-11-28 & 8 \\
Physics Reports&	2003-08-08 & 2 \\
Progress in Particle and Nuclear Physics&	2003-08-29 & 2 \\
Publications of the Astronomical Society of Australia&	2004-07-30 & 9 \\
Publications of the Astronomical Society of Japan&	2007-04-06 & 6 \\
Reviews of Modern Physics&	2003-08-29 & 1 \\
Science&	2003-07-18 & 26 \\
\hline\hline
\end{tabular}
\end{center}
\end{table}

\end{document}